\documentclass[pra,aps,twocolumn]{revtex4-1}
\usepackage{bm,dcolumn,amsmath,graphicx,amsfonts,amssymb,lipsum,mathtools}
\usepackage{epsfig}
\usepackage{tikz}
\usepackage{xcolor}

\begin{document}

\title{Atomic clocks highly sensitive to the variation of the fine structure constant based on 
 Hf II, Hf IV, and W VI ions}

	\author{Saleh O. Allehabi}
	\author{V. A. Dzuba}
	\author{V. V. Flambaum}

	\affiliation{School of Physics, University of New South Wales, Sydney 2052, Australia}
	
	\date{\today}
	
\begin{abstract}
We  demonstrate that several metastable  excited states in Hf~II, Hf~IV and W~VI ions may be good clock states since they are sufficiently long-living and are not sensitive to the perturbations. Cooling E1 transitions are available. Energy levels, Land\'{e} $g$-factors, transition amplitudes for  electric dipole (E1), electric quadrupole (E2), and magnetic dipole (M1) transitions, lifetimes, and electric quadrupole moments for Hf~II, Hf~IV, and W~VI ions are investigated using  a combination of  several methods of relativistic many-body calculations including   the configuration interaction (CI), linearized coupled-cluster single-doubles (SD) and many-body perturbation theory (CI+SD),  and  also the configuration interaction with perturbation theory (CIPT).  Scalar polarizabilities of the ground states and the clock states have been  calculated to determine the black body radiation (BBR) shifts. We have found  that the relative BBR shifts for these transitions range between 10$^{-16}$ $-$ 10$^{-18}$.  A linear combination of two clock transition frequencies allows one to further suppress BBR.
Several $5d$ - $6s$ single-electron clock transitions  ensure high sensitivity of the transition frequencies to the variation of the fine structure constant $\alpha$ and may be used to search for dark matter producing this variation of $\alpha$. The enhancement coefficient for $\alpha$  variation reaches $K=8.3$. Six stable isotopes  of Hf and 5 stable isotopes in W allow one to make King plots and search for new interactions  mediated by scalar particles or other mechanisms. 


	\end{abstract}
	
	\maketitle
	
\section{Introduction}
	
Atomic clocks possess high degree of accuracy
allowing them to be used for a wide variety of scientific and industrial applications.  In recent years,  optical lattice atomic clock and the ion clock both have been significantly enhanced to achieve uncertainties or stabilities of 10$^{-18} $~\cite{Ref4,Ref2,Ref6,Ref7,Freq.4,Freq.2,Freq.5,Ref9}. 

Due to the high accuracy of the frequency measurement of optical clock transitions, these transitions can be used not only to ensure timekeeping but also to search for new physics, such as local Lorentz invariance violation (LLI), 
 time-variation of the fundamental constants ($\alpha=e^2/\hbar c$) and other phenomena which go beyond the Standard Model (see, e.g., Refs.~\cite{Deltaj1,Deltaj2,Ref1,Ref2,Ref3,Ref4,Ref5,Ref6,Ref7,Ref8}). 

Most of operating optical clocks use the $^1$S$_0$ to $^3$P$_0$ transition between states of the $ns_{1/2}^2$ and $ns_{1/2}np_{1/2}$ configurations.
These transitions have low sensitivity  to  variation of the fine structure constant~\cite{two-el,CJP}. It was shown in Ref.~\cite{Deltaj1,Deltaj2} that maximum sensitivity to the $\alpha$-variation corresponds to the maximum change in the total angular momentum $j$ of the equivalent singe-electron transitions.
However, the above mentioned transitions are the $ns_{1/2}$ to $np_{1/2}$ transitions with $\Delta j$=0. It was suggested in Ref.~\cite{Deltaj1,Deltaj2,sd} to use transitions between states of different configurations. The most prominent example of this kind among operating optical clocks is the clock based on Yb$^+$ ion, in which the $4f_{7/2}$ - $6s_{1/2}$ and $6s_{1/2}$ - $5d_{5/2}$ transitions are used for time keeping and constraining of the time-variation of the fine structure constant~\cite{YbII-IIIa,YbII-IIIb,YbII-IIIexp}. A number of the promising transitions were studied in earlier works~\cite{sd,YbS,CuAgAu,CuYb}.
In the present paper we continue the search for promising candidates and study the $s-d$ and $d_{3/2}-d_{5/2}$ clock transitions in Hf~II, Hf~IV, and W~VI.
An important advantage of these systems is existence of sufficiently large number of stable isotopes of Hf and W. Hf has six stable isotopes, including four isotopes with zero nuclear spin (this includes a long-living $^{174}$Hf isotope with lifetime of $\sim 2\times 10^{15}$~years and natural abundance 0.16\%). W has five stable isotopes with three zero nuclear spin isotopes. This allows the use of the isotopes in search for non-linearities of King plot and study of possible sources of it, including new interactions  mediated by scalar particles, nuclear structure, etc. In addition, Hf~II ion has three metastable states, making it possible to construct two independent combinations of the frequencies of the clock transitions with suppressed black body radiation shift. Measuring one such combined frequency against the other over long period of time is a highly sensitive tool for the search of the time variation of the fine structure constant.

We provide a detailed analysis of the electronic characteristics of certain low-lying states of these systems.
We use the CI+SD (configuration interaction with  singes-doubles coupled cluster~\cite{Dzuba_SD+CI}) and the CIPT (configuration interaction with perturbation theory~\cite{CIPT}) methods for our calculations. Our studies investigate the energy levels, Land\'{e} $g$-factors, transition amplitudes, E1, M1, and E2 transitions for the low-lying states, lifetimes, and quadrupole moments. Using the technique described in Ref.~\cite{symmetry}, we also calculate the scalar polarizabilities of the ground states and the clock states in order to determine the black body radiation (BBR) shifts. The sensitivity to the variation of the fine structure constant is estimated by calculating excitation energies with different values of $\alpha$ in the computer codes. 
We demonstrate that the considered clocks  
are good candidates for very accurate timekeeping and are sensitive to new physics.

\section{Method of calculation}
	\subsection{Calculation of energy levels}
The Hf~II, Hf~IV, W~VI ions  
have similar electron structure with the [$1s^{2},..., 5s^2 5p^6 4f^{14}$ ] closed-shell core and three valence electrons in Hf~II and one valence electron in Hf~IV and W~VI. The calculations are performed by combining the configuration interaction (CI) technique with the linearized single-double coupled cluster (SD) method, as described in Ref.~\cite{Dzuba_SD+CI}. 
We start with the relativistic Hartree-Fock (RHF) calculations for the closed-shell core, which corresponds to the $V^{N-M}$ approximation~\cite{Dzuba1}.
Here $N$ is the total number of electrons in an atom or ion, and $M$ is the number of valence electrons ($M$= 3 for Hf~II and $M$= 1 for Hf~IV and W~VI). The RHF Hamiltonian has the form: 
\begin{equation} \label{e:RHF}
\hat H^{\rm RHF}= c\bm{\alpha}\cdot\bm{p}+(\beta -1)mc^2+V_{\rm nuc}(r)+V_{\rm core}(r),
\end{equation}
where $c$ is the speed of light, $\bm{\alpha}$ and $\beta$ are the Dirac matrixes, $\bm{p}$ is the electron momentum, $m$ is the electron mass, $V_{\rm nuc}$ is the nuclear potential obtained by integrating Fermi distribution of nuclear charge density, $V_{\rm core}(r)$ is the self-consistent RHF potential created by the electrons of the closed-shell core. 

The B-spline method is used to construct the set of single-electron basis states~\cite{Johnson_Bspline,Johnson_Bspline2}. The states are defined as linear combinations of B-splines which are eigenstates of the RHF Hamiltonian (\ref{e:RHF}). 40 B-splines of the order 9 are calculated within a box of radius  $R_{\rm max}$ = 40$a_B$ and an orbital angular momentum of 0~$\leq$~\textit{l}~$\leq$~6. The basis states are used for solving the linearized single-double couple-cluster (SD) equations and for generating the many-electron states for configuration interaction (CI) calculations. The correlation operators $\Sigma_1$ and $\Sigma_2$~\cite{Dzuba_SD+CI,CPM,Sigma2} are obtained by solving the SD equations for the core and then for the valence states.  The one-electron $\Sigma_1$ operator represents the correlation interaction between valence electrons and electrons in the core \cite{CPM}. The two-electron $\Sigma_2$ operator is interpreted as the screening of Coulomb interaction between valence electrons by core electrons \cite{Sigma2}. 

The CI Hamiltonian with the $\Sigma_1$ and $\Sigma_2$ operators included is 
\begin{equation}\label{e:HCI}
\hat{H}^{\mathrm{eff}}=\sum_{i=1}^{M}\left(\hat{H}^{\mathrm{RHF}}+\Sigma_{1}\right)_{i}+\sum_{i<j}^{M}\left(\frac{e^{2}}{\left|r_{i}-r_{j}\right|}+\Sigma_{2 i j}\right).
\end{equation}
Here summation goes over valance electrons, $i$ and $j$ numerate valence electrons, and $e$ is the electron charge. 
The size of the CI matrix is huge if the number of valence electrons is large (M $\ge$ 3). 
In calculations for Hf~II we use the CIPT technique~\cite{CIPT} for dramatic increase of the efficiency of the calculations at the cost of very little sacrifice in the accuracy of the results. This is achieved by treating highly excited states perturbatively, which allows to reduce the size of the effective CI matrix by several orders of magnitude (see Ref.~\cite{CIPT} for details).




The Land\'{e} $g$-factors for low-laying states are investigated in all systems. These factors are calculated as expectation values of the magnetic dipole (M1) operator and are compared with the non-relativistic expression
\begin{equation}
	g(J,L,S) = 1 + \frac{J(J+1)-L(L+1)+S(S+1)}{2J(J+1)},
	\label{e:g}
\end{equation}
where $S$ is total spin and $L$ is total angular momentum for the valence electrons, $J$ is corresponding total momentum ($\mathbf{J} = \mathbf{L} + \mathbf{S}$). This comparison of $g$-factors helps in the level identification.


\subsection{Calculation of transition amplitudes and lifetimes}
	
The time-dependent Hartree-Fock (TDHF) method (which corresponds to the well-known random phase approximation (RPA))
 is used to compute transition amplitudes.  The RPA equations for the core can be written as
\begin{equation}\label{e:RPA}
	\left(\hat H^{\rm RHF}-\epsilon_c\right)\delta\psi_c=-\left(\hat f+\delta V^{f}_{\rm core}\right)\psi_c
\end{equation}
The operator $\hat f$  refers to an external field. The index $c$ denotes single-electron states in the core, $\psi_c$ is a single electron wave function, $\delta\psi_c$ is a correction to the state $c$ due to an external field, $\delta V^{f}_{\rm core}$ is the correction to the self-consistent RHF potential caused by the change of all core states in the external field  (see e.g. Ref. ~\cite{CPM}).  The RPA equations (\ref{e:RPA}) are solved self-consistently for all states in the core. The transition amplitudes are found by calculating matrix elements between states $a$ and $b$ by the formula 
\begin{equation}\label{e:Ab}
	A_{a b}=\left\langle b\left|\hat f+\delta V^{f}_{\rm core }\right| a\right\rangle
\end{equation}
Here, $|a\rangle$ and $|b\rangle$ are the many-electron wave functions calculated with the CI method described above. 

In this study, electric dipole (E1), electric quadrupole (E2), and magnetic dipole (M1) rates are taken into account, and they are calculated according to the following equations (in atomic units):
\begin{equation}\label{e:Td}
	(T_{ab})_{E1,M1} = \frac{4}{3}(\alpha\omega)^3 \frac{(A^2_{ab})_{E1,M1}}{2J_b+1},
\end{equation}

\begin{equation}\label{e:Tq2}
(T_{ab})_{E2} = \frac{1}{15}(\alpha\omega)^5 \frac{(A^2_{ab})_{E2}}{2J_b+1}.
\end{equation}
Here $\alpha$ is the fine structure constant ($\alpha\approx\frac{1}{137}$), $\omega_{ab}$ is the frequency of the transition, $A_{ab}$ is the transition amplitude (\ref{e:Ab}), and $J_b$ is the total angular momentum of the upper state $b$. Note that magnetic amplitudes $(A_{ab})_{M1}$ contain the Bohr magneton $\mu_B$ ($\mu_B = \alpha/2 \approx 3.65 \times 10^{-3}$ in atomic units). 

The lifetimes $\tau_b$ of each excited state $b$, expressed in seconds, can be found as follows:
\begin{equation}\label{e:tau}
	\tau_b =  2.41\times 10^{-17}\bigg/\sum_a T_{ab}
\end{equation}
where the summation goes over all possible transitions to lower states $a$.

		\begin{table*}
		
		\caption{\label{t:Energy}
			Excitation energies ($E$), Land\'{e} \textit{g}-factors, and lifetimes ($\tau$) for the first excited states of Hf~II, Hf~IV and W~VI. Possible clock states are indicated by bold state numbers. Odd states can be used for cooling.} 
		\begin{ruledtabular}
			\begin{tabular}{cc cc cc cc cc}
				&&&&
			
					\multicolumn{2}{c}{E [cm$^{-1}$]}&
						\multicolumn{2}{c}{\textit{g}-factor}&

				\multicolumn{2}{c}{$\tau$}\\
				\cline{5-6}
					\cline{7-8}
						\cline{9-10}
				\multicolumn{1}{c}{No.}& 
				\multicolumn{1}{c}{Conf.}&
				\multicolumn{1}{c}{Term}&

				\multicolumn{1}{c}{$J$}&
				\multicolumn{1}{c}{Present}&
				\multicolumn{1}{l}{Expt.}&
								\multicolumn{1}{c}{Present}&
				\multicolumn{1}{l}{NIST~\cite{NIST}}&
					\multicolumn{1}{c}{Present}&
						\multicolumn{1}{c}{Ref. }\\
				\hline

				\multicolumn{5}{c}{\textbf{Hf~II}}&
				\multicolumn{1}{c}{\cite{NIST}}&&&&
					\multicolumn{1}{c}{\cite{Tran}$ _{\rm Expt. }$}\\

	1 & $5d6s^2$& $^2${D}& {3/2} &0 &0&0.793&0.787&&\\
{\bf 2} & $5d6s^2$& $^2${D}& {5/2} &3054&3050.88&1.175&1.173&3.23 s&\\
{\bf 3}& $5d^{2}6s$& $^4${F}& {3/2} &3578&3644.65&0.415&0.425&66.6 s&\\
{\bf 4}& $5d^{2}6s$& $^4${F}& {5/2} &4312&4904.85&1.055&1.052&9.7 s&\\
5& $5d^{2}6s$& $^4${P}& {1/2} &11675&11951.70&2.653&2.598&&\\
6& $5d^{2}6s$& $^2${F}& {5/2} &11783&12070.46&0.901&0.964&&\\
7& $5d^{2}6s$& $^4${P}& {3/2} &11781&12920.94&1.694&1.664&&\\
8& $5d^{2}6s$& $^4${P}& {5/2} &12581&13485.56&1.467&1.410&&\\

9& $5d^{2}6s$& $^2${D}& {3/2} &13836&14359.42&1.075&1.034&&\\

10& $5d^{2}6s$& $^2${P}& {1/2} &13995&15254.29&0.690&0.737&&\\

11& $5d^{2}6s$& $^2${D}& {5/2} &17352&17368.87 &1.200&1.273 &&\\

12& $5d^{2}6s$& $^2${P}& {3/2} &17199& 17830.34 &0.670&1.122&&\\

13& $5d^{3}$& $^4${F}& {3/2} &18528&18897.64&0.839&0.446 &&\\

14& $5d^{3}$& $^4${F}& {5/2} &18284&20134.94&1.118&1.030 &&\\

15& $5d^{3}$& $^4${P}& {1/2} &24773&26996.51&2.610&2.58&&\\

16& $5d^{3}$& $^4${P}& {3/2} &25797&27285.13&1.697&1.643&&\\


17& $5d6s6p$& $^4${F}$\rm ^o$& {3/2} &28580&28068.79&0.516&0.512&40.3 ns&39.4$\pm$ 0.2 ns\\


			\hline

\multicolumn{5}{c}{\textbf{Hf~IV}}&
\multicolumn{1}{c}{\cite{Klinkenberg}}&&&&
\multicolumn{1}{c}{}\\

1 & $5d$& $^2$D&{3/2} &0&0&0.800&&&\\
{\bf 2} & $5d$& $^2${D}&{5/2} &4721 &4692&1.200&&0.90 s&\\
{\bf 3} &$6s$&$^2${S}&{1/2} &17530&18380&2.000&&0.321 s&\\
4 & $6p$& $^2${P}$\rm ^o$&{1/2} &66611&67039&0.667&&0.78 ns&\\

	\hline
\multicolumn{5}{c}{\textbf{W~VI}}&
\multicolumn{1}{c}{\cite{NIST}}&&&&
\multicolumn{1}{c}{\cite{W5+}$ _{\rm Theo. }$}\\

1 & $5d$& $^2$D&{3/2} &0&0&0.800&&&\\
{\bf 2} & $5d$& $^2${D}&{5/2} &8726 &8709.3&1.200&&0.14 s& 0.14 s\\
3 &$6s$&$^2${S}&{1/2} &78316&79431.3&2.000&&&\\
4 & $6p$& $^2${P}$\rm ^o$&{1/2} &146912&147553.1&0.667&& 0.18 ns&0.184 ns\\

			\end{tabular}
		\end{ruledtabular}
	
	\end{table*}

\section{Results}
\subsection{Energy levels, Land\'{e} $g$-factors, Transition amplitudes, and Lifetimes}
	

The results for energy levels, $g$-factors and lifetimes of low-lying states of Hf~II, Hf~IV and W~VI are presented in Table~\ref{t:Energy} and compared with available experimental data.
The data in the table indicate excellent agreement between theory and experiment.
In most states, the deviations of the calculated energies from the observed values are within 1000~cm$ ^{-1} $. 
The agreement is also good between calculated and experimental $g$-factors of Hf~II, where experimental data are available. This is important for correct identification of the states. One noticeable exemption refers to states 12 and 13 where the difference between theory and experiment is significant. These states have the same parity and total momentum $J$, the energy interval between them is small ($\sim$ 1000~cm$^{-1}$). This means that the states are strongly mixed. Note that the sums of the theoretical and experimental $g$-factors of these states are very close. This indicates that the two-level mixing approximation works very well for this pair of states. In principle, mixing coefficients can be corrected using experimental $g$-factors. See the discussion of the sensitivity of the clock states to the variation of the fine structure constant (subsection \ref{s:alpha}).

	
Our results for transition amplitudes and transition probabilities, together with experimental data and earlier calculated results, where available, are shown in Table~\ref{t:Tran}. We consider only those low-lying states which are connected to clock or cooling states through electric dipole (E1), magnetic dipole (M1), or electric quadrupole (E2) transitions. Comparing our results on the transition rates with those from previous studies, we find good agreement. Note that experimental values of the  frequencies from the NIST database  have been used to calculate transition probabilities.
	
Based on the transition rates displayed in Table \ref{t:Tran}, we derived the lifetimes of the excited states (clock and cooling states) of all the atomic systems using Eq.~(\ref{e:tau}) and presented them in Table \ref{t:Energy}. The lifetimes of the states presented in the table were calculated with taking into account all possible transitions to lower states. The results show consistency with previous studies.
		
\begin{table*}
	
	\caption{\label{t:Tran} Transition amplitudes (\textit{A}, a.u.) and transition probabilities (T, 1/s) evaluated with NIST frequencies for some low  states. 5.67[-3] means $5.67 \times 10^{-3}$, etc. }
	
	\begin{ruledtabular}
		\begin{tabular}{ll ll rl c}
			&&
			\multicolumn{2}{c}{($\omega$), NIST \cite{NIST} }&
			\multicolumn{2}{c}{Present}&
				\multicolumn{1}{c}{Ref.}\\
			
			\cline{3-4}
			\cline{5-6}

			\multicolumn{1}{c}{Transition}& 
			\multicolumn{1}{c}{Type}&
			\multicolumn{1}{c}{ [cm$^{-1}$]}&
			\multicolumn{1}{c}{ [a.u.]}&

			\multicolumn{1}{c}{\textit{A} [a.u]}&
			\multicolumn{1}{c}{T [s$^{-1}$]}&
			\multicolumn{1}{c}{T [s$^{-1}$]}\\
			
			\hline
			
				 \multicolumn{6}{c}{\textbf{Hf~II}}&
				 					\multicolumn{1}{c}{\cite{Tran}$ _{\rm Expt. }$}\\
			2 $-$ 1       & M1   & 3050.88  & 0.0139      & 5.67[-3]  & 0.309 &   \\
			2 $-$ 1      & E2   & 3050.88  & 0.0139      & -0.309 & 4.697[-7]&  \\
			3 $-$ 1      & M1   & 3644.65  & 0.0166      & -7.49[-4] & 1.375[-2]&  \\
			3 $-$ 1      & E2   & 3644.65  & 0.0166      & 0.265  & 1.259[-6] & \\
			3 $-$ 2      & M1   & 593.7    & 0.0027      & 3.53[-3]  & 1.318[-3]&  \\
			3 $-$ 2      & E2   & 593.7    & 0.0027      & 0.341  & 0.241[-9]& \\
			4 $-$ 1      & M1   & 4904.85  & 0.0223      & -7.54[-4] & 2.261[-2]&  \\
			4 $-$ 1      & E2   & 4904.85  & 0.0223      & -0.420 & 9.325[-6]&  \\
			4 $-$ 2     & M1   & 1853.97  & 0.0084      & -1.63[-3] & 5.743[-3]&  \\
			4 $-$ 2     & E2   & 1853.97  & 0.0084      & -0.259 & 2.734[-8]&  \\
			4 $-$ 3     & M1   & 1260.2   & 0.0057      & 1.05[-2]  & 7.497[-2]&  \\
			4 $-$ 3     & E2   & 1260.2   & 0.0057      & 1.56     & 1.439[-7]&  \\

		17 $-$ 1	&E1 & 28068.79 & 0.1279 & 1.200  & 16.11[6]&17.6[6] $\pm$ 0.9 \\
		17 $-$ 2	&E1 & 25017.91 & 0.1140 & -0.068 & 3.674[4]& \\
		17 $-$ 3	&E1 & 24424.14 & 0.1113 & 1.024  & 7.728[6]& 7.0[6] $\pm$ 0.4 \\
		17 $-$ 4	&E1 & 23163.94 & 0.1055 & 0.016  & 1.660[3]& \\
		17 $-$ 5	&E1 & 16117.09 & 0.0734 & 0.068  & 9.712[3]&2.1[4] $\pm$ 0.003 \\
	    17 $-$ 6		&E1 & 15998.33 & 0.0729 & 0.522  & 0.565[6]&0.50[6] $\pm$ 0.09 \\
		17 $-$ 7 	&E1 & 15147.85 & 0.0690 & 0.110  & 0.021[6]& \\
		17 $-$ 8	&E1 & 14583.23 & 0.0664 & 0.128  & 0.026[6]&0.060[6] $\pm$ 0.011 \\

	17 $-$ 9		&E1 & 13709.37 & 0.0625 & 0.030  & 0.012[5]& \\
	17 $-$ 10		&E1 & 12814.5  & 0.0584 & -0.301 & 0.096[6]&0.054[6] $\pm$ 0.011 \\
	17 $-$ 11		&E1 & 10699.92 & 0.0488 & 0.224  & 0.031[6]& \\

	17 $-$ 12		&E1 & 10238.45 & 0.0466 & 0.517  & 0.145[6]& \\
	17 $-$ 13		&E1 & 9171.15  & 0.0418 & 0.342  & 0.046[6]&0.081[6] $\pm$ 0.019 \\
	17 $-$ 14		&E1 & 7933.85  & 0.0361 & -0.265 & 0.018[6]& \\
	17 $-$ 15		&E1 & 1072.28  & 0.0049 & -0.069 & 2.964& \\
	17 $-$ 16		&E1 & 783.66   & 0.0036 & 0.023  & 0.127&\\
	\hline
	
		 \multicolumn{6}{c}{\textbf{Hf~IV}}\\
		 
		 	2 $-$ 1       & M1   & 4692	&0.0214	&-5.66[-3]	&1.115 &   \\
		 2 $-$ 1      & E2   & 4692	&0.0214	&-2.43	&2.501[-4]&  \\
		 3 $-$ 1      & M1   & 18380	&0.0837	&2.22[-6]	&3.091[-5]&  \\
		 		 3 $-$ 1      &E2  & 18380	&0.0837	&4.40&	2.267&  \\
		 		 		 3 $-$ 2      & E2   & 13688&0.0624&	-5.62&	0.850&  \\
		 		 		 
		 		 		 	4 $-$ 1	&E1 & 67039	&0.3055	&1.62&	7.996[8]& \\
		 		 		 	
		 		 		 	4 $-$ 3	&E1 & 48659	&0.2217&	2.04&	4.860[8]& \\
		 		 		 	\hline
		 		 		 	
		 		 		 			 \multicolumn{6}{c}{\textbf{W~VI}}&
\multicolumn{1}{c}{\cite{W5+}$ _{\rm Theo. }$}\\
		 		 		 	
		 		 		 	2 $-$ 1       & M1   & 8709.3	&0.0397&	-5.65[-3]&	7.126 &7.12   \\
		 		 		 	2 $-$ 1      & E2   & 8709.3	&0.0397&-1.60&	2.400[-3]& 2.54[-3] \\	
		 		 		 	4 $-$ 1	&E1 & 147553.1	&0.6723&	1.18&	4.529[9]& \\
		 		 		 	
		 		 		 	4 $-$ 3	&E1 & 68121.8&	0.3104&	1.70&	9.218[8]& \\
		\end{tabular}
	\end{ruledtabular}
\end{table*}
	
\subsection{Polarizabilities and blackbody radiation shifts}

Scalar polarizability is one of the key properties of atoms that sets their chemical characteristics. For establishing optical clocks, the values of the static and dynamic scalar polarizabilities should be taken into account. Scalar polarizabilities provide the value of the black body radiation (BBR) shift of the clock state frequency, which is a primary source of uncertainty for a clock.

The scalar polarizability $\alpha_{0v}$ of an atomic system in state $v$ can be expressed as a sum over a complete set of states $n$
\begin{equation}\label{e:pol}
\alpha_{0v}=\dfrac{2}{3(2J_v+1)}\sum_{n}\frac{A_{vn}^2}{E_n-E_v}.
\end{equation}
Here $J_v$ is the total angular momentum of state $v$, and $A_{vn}$ is the electric dipole transition amplitude (reduced matrix element). Notations $v$ and $n$ refer to the many-electron atomic states. It is convenient to present the polarizability as a sum of two terms, the polarizability of the closed-shell core and the contribution from the valence electrons. The polarizability of the core is given in the RPA approximation by 
\begin{equation}\label{e:pol0}
\alpha_{0\rm core}=\dfrac{2}{3}\sum_{cn}\frac{\langle c ||\hat d + \delta V_{\rm core}^d ||n\rangle \langle n ||\hat d ||c\rangle}{E_n-E_c}.
\end{equation}
Here summation over $c$  goes over core states, summation over $n$ is over a complete set of single-electron basis states, $\hat d =-er$ is the electric dipole (E1) operator in the length form, $\delta V_{\rm core}^d$ is the core polarisation correction to the E1 operator (see eq.(\ref{e:Ab})). Note that the RPA correction goes only into one of the two reduced matrix elements in (\ref{e:pol0})~\cite{pol0}.

For the calculation of the valence contribution to the polarizabilities of the ground and clock states of Hf~II we apply the technique developed in Ref.~\cite{symmetry} for atoms or ions with open shells. The method relies on Eq.~(\ref{e:pol}) and the Dalgarno-Lewis approach~\cite{Dalgarno:1955}, which reduces the summation in (\ref{e:pol}) to the solving of the matrix equation (see Ref.~\cite{Dalgarno:1955,symmetry} for more details). 
For Hf~IV and W~VI, which both have only one external electron above closed shells, we use direct summation in Eq. (\ref{e:pol}) over complete set of single-electron basis states. 

There is also a core-valence contribution to the polarizabilities which comes from the fact that calculation of the core polarizabilities is affected by valence electrons via Pauli blocking. We include this contribution by omitting in summation over $n$ in (\ref{e:pol0}) states occupied by valence electrons. 

The present results for the polarizabilities of the ground states and clock states for all considered atomic systems are shown in Table~\ref{t:pol}. 
According to the calculations, clock states of all atomic systems have polarizabilities similar to that of their ground states, with the notable exception of the Hf~IV third excited state, where this difference is approximately 14~$a_b^3$. This is because of the difference in the electronic configurations. The external electron in the ground state is in the $5d_{3/2}$ state while in clock state it is in the $6s_{1/2}$ state.

By using the values of scalar polarizability, we can figure out the BBR shift of a clock state at 300 K. The BBR shift in Hz is determined by the following expression (see e.g. ~\cite{BBR})
\begin{equation}\label{e:BBR}
\delta \omega_{\rm BBR} = -8.611 \times  10^{-3}  \left(\frac{T}{300~K}\right)^4 \Delta \alpha_0,
\end{equation}
where $T$ is a temperature in K (e.g., room-temperature $T$= 300 K), $ \Delta \alpha_0= \alpha_0({\rm CS}) - \alpha_0({\rm GS})$, is the difference between the clock state and the ground-state polarizabilities presented in atomic units.
The  BBR shifts  for clock states investigated in this paper are presented in Table~\ref{t:pol}.  
The relative BBR shifts in the 2-1 transition in Hf~IV and 2-1 transition in W~VI are among the smallest considered so far,  they are $4.3\times 10^{-18}$ and $ 2.3\times 10^{-18}$ respectively, while  BBR shifts in other transitions are $\sim 10^{-16}$, similar to BBR shifts in other atomic clocks, see e.g.  ~\cite{BBR1, BBR2,BBR3,BBR4,BBR5}. A linear combination of two clock transition frequencies allows one to cancel BBR shifts \cite{CancelBBR}.

		\begin{table*}
		\caption{\label{t:pol}
			Scalar static polarizabilities of the ground states, $\alpha_0({\rm GS})$, and clock states, $\alpha_0({\rm CS})$,  and BBR frequency shifts for the clock transition.  $\delta\omega_{BBR}$/$\omega$ is the fractional contribution of the BBR shift; where $\omega$ is the clock transition frequency. Total means total scalar polarizability (core + valence). 
		}
		\begin{ruledtabular}
			\begin{tabular}{ccccccclllr}
				
				\multicolumn{1}{c}{Clock}&
				\multicolumn{3}{c}{$\alpha_0({\rm GS})$[$a_B^3$]}&
				\multicolumn{3}{c}{$\alpha_0({\rm CS})$[$a_B^3$]}&&
				\multicolumn{3}{c}{BBR, (\textit{T}= 300 K)} \\

				\cline{2-4}
				\cline{5-7}
				\cline{9-11}
				\multicolumn{1}{c}{transition}&

				\multicolumn{1}{c}{Core}&
				\multicolumn{1}{c}{Valence} &
				\multicolumn{1}{c}{Total} &
				
				\multicolumn{1}{c}{Core}&
				\multicolumn{1}{c}{Valence} &
				\multicolumn{1}{c}{Total} &
				\multicolumn{1}{c}{$\Delta \alpha_0$} &
				\multicolumn{1}{c}{$\delta\omega_{BBR}$[Hz]}&
				\multicolumn{1}{c}{$\omega$[Hz]}&
				\multicolumn{1}{c}{$\delta\omega_{BBR}$/$\omega$} \\
				\hline
				
			 \multicolumn{11}{c}{\textbf{Hf~II}}\\
				
2 $-$ 1 & 2.72 & 48.04 & 50.76 &   2.72 & 43.22 & 45.94 & -4.93 & 0.043 & 1.093[14] & 3.9[-16] \\			
3 $-$ 1 & 2.72 & 48.04 & 50.76 &   2.61 & 40.10 & 42.71 & -8.05 & 0.070 & 1.093[14] & 6.4[-16] \\			
4 $-$ 1 &2.72 & 48.04 & 50.76 & 2.61 & 42.84 & 45.45 & -5.31 & 0.046 & 1.470[14] & 3.2[-16] \\
\\

	 \multicolumn{11}{c}{\textbf{Hf~IV}}\\
	 
2 $-$ 1 & 3.06 & 2.86 & 5.92 & 2.98 & 2.86 & 5.85 & -0.07 & 0.60[-3] & 1.404[14] & 4.3[-18] \\	 
3 $-$ 1 & 3.06 & 2.86 &  5.92 & 3.13 & 16.84 & 19.97 & 14.05 & -0.12 & 5.510[14] & -2.2[-16] \\
	 \\
	 
	 	 \multicolumn{11}{c}{\textbf{	W~VI}}\\
	 
2 $-$ 1 & 1.96 & 0.76 &  2.73 & 1.92 & 0.75 & 2.66 & -0.07 & 0.60[-3] & 2.611[14] & 2.3[-18] \\	 

		\end{tabular}
	\end{ruledtabular}
\end{table*}

\subsection{Electric quadrupole moments}

\begin{table}[!]
	\caption{\label{t:Q}
			Quadrupole moment ($\Theta$, a.u.) of the ground state and the considered optical clock states.} 		
		\begin{ruledtabular}
			\begin{tabular}{crccccc}
				\multicolumn{1}{c}{No.}& 
				\multicolumn{1}{c}{Conf.}&
				\multicolumn{1}{c}{Term}&
				\multicolumn{1}{c}{$J$}&
				\multicolumn{1}{c}{$E$ (cm$^{-1}$)}&
				\multicolumn{1}{c}{ ME (a.u.) }&
				\multicolumn{1}{c}{ $\Theta$ }\\&&&&&
				\multicolumn{1}{c}{ $\left\langle J\|\hat\Theta\| J\right\rangle$}\\
				\hline
				\multicolumn{7}{c}{\textbf{Hf~II}}\\
				1& $5d6s^2$& $^2${D}& {3/2}&0&-2.910[-2]&-6.507[-3]\\
				2& $5d6s^2$& $^2${D}& {5/2}&3050.88&-5.806[-1]& -1.415[-1]\\
				3& $5d^{2}6s$& $^4${F}& {3/2}&3644.65&-1.797&-0.401\\
				4& $5d^{2}6s$& $^4${F}& {5/2}&4904.85&-1.621&-0.395\\
				\\
				\multicolumn{7}{c}{\textbf{Hf~IV}}\\
				1 & $5d$& $^2${D}&{3/2} &0&-3.608&-0.807\\
				2 & $5d$& $^2${D}&{5/2} &4692&-4.956&-1.210\\
				\\
				\multicolumn{7}{c}{\textbf{W~VI}}\\
				1 & $5d$& $^2${D}&{3/2} &0&-2.381&-0.532\\
				2 & $5d$& $^2${D}&{5/2} &8709.3&-3.263&-0.796\\
				\end{tabular}
		\end{ruledtabular}
	\end{table}

As it was discussed above, the search for clock transitions sensitive to the variation of the fine structure constant leads us to transitions with large change change of the total momentum of the equivalent singe-electron transitions. As a consequence, at least one of the states may have relatively large value of the total momentum $J$ (e.g., $J > 1$). This means that the state is sensitive to
the electric quadrupole shift. Therefore, it is important to know the value of this shift.
The corresponding term in the Hamiltonian is (see, e.g. \cite{Q})


\begin{equation}
H_{\mathcal{Q}}=\sum_{q=1}^{-1}(-1)^{q} \nabla \mathcal{E}_{q}^{(2)} \hat\Theta_{-q}
\end{equation}

The tensor $ \nabla \mathcal{E}_{q}^{(2)}$ is the external electric field gradient at the position of the system, $\Theta_{q}$ describes the electric-quadrupole operator, $\hat\Theta_{q}= |e|r^2C_q^{(2)}$, the same as for E2 transitions.


The electric quadrupole moments $\Theta$ is defined as the expectation value of $\Theta_{0}$ for the extended state
\begin{equation}
\begin{aligned}
\Theta &=\left\langle n J J\left|\hat\Theta_{0}\right| n J J\right\rangle\\
&\equiv\left\langle J\|\hat\Theta\| J\right\rangle \sqrt{\frac{J\left(2 J-1\right)}{\left(2 J+3\right)\left(2 J+1\right)\left(J+1\right)}},
\end{aligned}
\end{equation}
where $\left\langle J\|\hat\Theta\| J\right\rangle$ is the reduced ME of the electric quadrupole operator. We compute the values of $\Theta$ using the CI+SD and RPA methods described in the previous section. 

In Table~\ref{t:Q} we display the reduced ME of the electric quadrupole operator and the quadrupole moment $\Theta$  values for all considered states. 
The quadrupole momentum of the ground state of Hf~II is anomalously small. This is due to the mixing between states of the $6s^25d$ and $6s5d^2$ configurations leading to strong cancelation between terms proportional to the $\langle 5d_{3/2} ||\hat\Theta|| 5d_{3/2}\rangle$ ME and terms proportional to the $\langle 6s_{1/2} ||\hat\Theta|| 5d_{3/2}\rangle$ ME. Strong cancellation is probably accidental, which means that the result is likely to be not very stable and may vary significantly under variation of the computation procedure. In states where the $5s5d^2$ configuration dominates the value of the quadrupole moment is not suppressed.  
The values of the quadrupole moments for both excited clock states are almost the same. In the case of the Hf~IV ion, the difference between the values in the ground state and in the excited state is much smaller than that in the Hf$ ^{+}$ ion. The quadrupole moment in the ground state is about 1.5 times larger  than that in the first excited state. This is because both states (the ground and the excited) have the same electron configuration. Note that the quadrupole moment of the second excited state ($6s$\ $^2${S}$_{1/2}$) is zero because the total angular momentum J is 1/2. The W~VI ion  has differences similar to that in the Hf~IV ion. 

It is worth to note that the quadrupole shift in odd isotopes of Hf~II and Hf~IV ($^{177}$Hf, $I=7/2$, and $^{179}$Hf, $I=9/2$) can be totally avoided  when working with specific hyperfine components of the states, namely substates with $F=3$ and $F_z=2$, since $\Delta E \sim 3F_z^2 - F(F+1)$. Such substates exist for both, ground and clock states, in both isotopes. Another way of suppressing the quadrupole shift is by averaging over transitions between different hyperfine or Zeeman components~\cite{YbS,Q}. Averaging over Zeeman components should work for even isotopes too.

\subsection{Sensitivity of the clock transitions to the variation of the fine structure constant}
\label{s:alpha}

\begin{table}[!]
	\caption{\label{t:q}
		Sensitivity of clock transitions to variation of the fine-structure constant ($q, K$) .} 
	\begin{ruledtabular}
		\begin{tabular}{crccccc}
			
			\multicolumn{1}{c}{No.}& 
			\multicolumn{1}{c}{Conf.}&
				\multicolumn{1}{c}{Term}&
					\multicolumn{1}{c}{$J$}&

			\multicolumn{1}{c}{$\omega$ (cm$^{-1}$)}&
			
			\multicolumn{1}{c}{$q$(cm$^{-1}$) }&
				\multicolumn{1}{c}{$K$}\\
			\hline
			 \multicolumn{7}{c}{\textbf{Hf~II}}\\
		2& $5d6s^{2}$& $^2${D}& {5/2}&3050.88&5631&3.65\\	
		 3& $5d^{2}6s$& $^4${F}& {3/2}&3644.65&15060&8.30\\
		 4& $5d^{2}6s$& $^4${F}& {5/2}&4904.85&15002&6.16\\
		 			\\
		 \multicolumn{7}{c}{\textbf{Hf~IV}}\\
		 2 & $5d$& $^2${D}&{5/2} &4692&4342&1.85\\
		 3 &$6s$&$^2${S}&{1/2} &18380&-24268&-2.64\\
		 			\\
\multicolumn{7}{c}{\textbf{W~VI}}\\
2 & $5d$& $^2${D}&{5/2} &8709.3&8609&1.98\\
		\end{tabular}
	\end{ruledtabular}
\end{table}

It has been shown that optical atomic clock transitions can be used to search for the time variation of the fine-structure constant $\alpha$ \cite{Deltaj1,Deltaj2,CJP}. The frequencies of these transitions depend differently on $\alpha$. By comparing the ratio of two clock frequencies over long periods of time, one can 
link any possible change in  the ratio of  frequencies to the time variation of $\alpha$. The ratio of frequencies  does not depend on the units one uses. 
In atomic units dependence of the optical transition frequencies appears due to the relativistic corrections  proportional to $\alpha^2$.   Therefore, we present  frequency  as 
 \begin{equation}\label{e:q}
 	\omega = \omega_0 + q\left[\left(\frac{\alpha}{\alpha_0}\right)^2-1\right],
 \end{equation}
 where $\alpha_0$ and $\omega_0$ are the present-day values of the fine structure constant and the frequency of the transition, $q$ is the sensitivity coefficient which comes from the atomic calculations~\cite{Deltaj1,Deltaj2,CJP}.
 The rate of the variation of $\omega_1/\omega_2$  is 
\begin{equation}\label{e:adot}
 	\frac{\dot \omega_1}{\omega_1} - \frac{\dot \omega_2}{\omega_2} = \left(K_1 - K_2 \right)\frac{\dot \alpha}{\alpha}.
\end{equation}
The dimensionless value $K=2q/\omega$ is often called the enhancement factor. We use the computer codes to calculate $q$ and $K$ by performing calculations of the frequencies with different values of $\alpha$ and calculating the derivative numerically as 
\begin{equation}
 	q=\frac{\omega(\delta)-\omega(-\delta)}{2\delta},
\end{equation}
 where $\delta =(\alpha/\alpha_0)^2-1$ (see Eq.~(\ref{e:q})). In order to achieve linear behaviour, the value of $\delta$  must be small; however, it must be large enough to suppress numerical noise. Most accurate results can be obtained by using $\delta =0.01$. The calculated values of $q$ and $K$ for all considered clock transitions  are summarised in Table~\ref{t:q}. We see that all values of the enhancement coefficients $K$ are significantly bigger than 1.
The enhancement factor for the 2-1 interval is 3.65. This is fine structure interval and in normal circumstances the interval $\propto (Z\alpha)^2$ and $K=2$. Here the factor is significantly larger due to strong mixing of the upper state with the states of the $5d^26s$ configuration.
Note that values of $K$ are positive, with only one negative $K$ factor for the $5d_{3/2} - 6s_{1/2}$ transition in the Hf~IV ion. Indeed, simple analytical estimate performed in Refs.~\cite{Deltaj1,Deltaj2} indicates that the transition from lower $j$ orbital in the ground state to higher $j$ orbital gives positive $K$ while  the transition from higher  $j$ orbital to lower  $j$ orbital gives negative $K$. 

\begin{table}[!]
\caption{\label{t:Kij}
Synthetic frequencies of Hf~II clock transitions and their sensitivity to variation of the fine-structure constant.
Indexes $i$ and $j$ correspond to the clock transitions from state number $i$ or $j$ (see Table~\ref{t:Energy}) to the ground state.} 
\begin{ruledtabular}
\begin{tabular}{cccrr}
\multicolumn{1}{c}{$i$}& 
\multicolumn{1}{c}{$j$}& 
\multicolumn{1}{c}{$\epsilon_{ij}$}& 
\multicolumn{1}{c}{$\omega_{ij}$ [cm$^{-1}$]}& 
\multicolumn{1}{c}{$K_{ij}$}\\
\hline 
  3 & 2 &  1.633  &   1336.86  &   -9.03 \\
  4 & 2 &  1.077  &   1618.83  &   11.26 \\
  4 & 3 &  0.660  &   2500.66  &   4.10 \\
\end{tabular}
\end{ruledtabular}
\end{table}

It was shown in Ref.~\cite{CancelBBR} that having two clock transitions in one atom or ion allows one to construct a "synthetic" frequency which is not sensitive to the BBR shift. Using such frequencies may benefit the search for the time variation of the fine structure constant. The Hf~II ion has three clock transitions. This means that one can contract two independent "synthetic" frequencies non-sensitive to the BBR shift. Measuring one such frequency against the other over long period of time allows highly sensitive search for the variation of the fine structure constant. Following Ref.~\cite{CancelBBR} we write a synthetic frequency as
\begin{equation}\label{e:syn}
\omega_{ij} = \omega_i - \epsilon_{ij} \omega_j,
\end{equation}
where $\epsilon_{ij} = \Delta \alpha_{0i}/\Delta \alpha_{0j}$. Since BBR shift $\propto \Delta \alpha_0$ (see Eq.(\ref{e:BBR})), the synthetic frequency (\ref{e:syn}) is not sensitive to it. If the fine structure constant $\alpha$ varies in time then the synthetic frequency varies as
\begin{equation}\label{e:dotnu}
\frac{\dot \omega_{ij}}{\omega_{ij}} = \frac{K_i\omega_i - \epsilon_{ij}K_j\omega_j}{\omega_i-\epsilon_{ij}\omega_j}\frac{\dot \alpha}{\alpha} \equiv K_{ij}\frac{\dot \alpha}{\alpha}.
\end{equation}
Table \ref{t:Kij} shows three possible synthetic frequencies for Hf~II. Any two of these frequencies can be used for searching of the time variation of the fine structure constant. For example, if the $\omega_{32}$ and $\omega_{42}$ frequencies are used then
\begin{equation}
\frac{\dot \omega_{32}}{\omega_{32}} - \frac{\dot \omega_{42}}{\omega_{42}} = -20 \frac{\dot \alpha}{\alpha}.
\end{equation}
The combinations of frequencies which are not sensitive to the BBR shift turn out to be very sensitive to the time variation of the fine structure constant.

 \section{Conclusions}

Metastable states of Hf~II, Hf~IV, and W~VI  ions are studied as candidates for high accuracy optical clocks which are highly sensitive to the variation of the fine structure constant $\alpha$. Slow drift and oscillating variation of $\alpha$ may be due to the interaction between scalar dark matter field  and electromagnetic field  \cite{TilburgBudker,StadnikPRL,StadnikPRA,HeesGuena}. Yukawa-type scalar field affecting $\alpha$ may also be produced by massive  bodies 
\cite{Leefer}. Transient  $\alpha$ variation may be produced  by passing of macroscopic forms of dark matter such as Bose stars and topological defects  \cite{DereviankoPospelov}. 
 
 Six stable isotopes  of Hf and five stable isotopes in W and several clock transitions in Hf and W ions allow one to make King plots and search for a new interaction  mediated by a scalar particle which leads to a non-linearity of the King plot \cite{Berengut,Geddes}.

Energy levels, lifetimes, transition rates, scalar polarizabilities of the clock and ground states, and BBR shifts have been calculated and the possibility for high accuracy of the time keeping has been  demonstrated. The studied transitions correspond to the $s-d$ or $d-s$ transitions between single-electron states.

The sensitivity coefficients to $\alpha$ variation $K$ have been  calculated and found to be among the highest compared to other operating or prospective atomic optical clocks. 
We found that constructing synthetic frequencies with suppressed sensitivity to the BBR shift leads to further increase in sensitivity to the variation of the fine structure constant.

 

\section{Acknowledgements}
This work was supported by the Australian Research Council Grants No. DP190100974 and DP200100150.  This research includes computations using the computational cluster Katana supported by Research Technology Services at UNSW Sydney~\cite{Katana}.

\end{document}